\documentclass[epjST]{svjour}
\usepackage{mhchem}
\usepackage{graphics}
\usepackage{color}
\usepackage{epsfig}

\def\gsim{\:\raisebox{-0.5ex}{$\stackrel{\textstyle>}{\sim}$}\:}

\def\bg#1{\mathchoice
{{\mbox{\boldmath $#1$}}}
{{\mbox{\boldmath $#1$}}}
{{\mbox{\boldmath $\scriptstyle #1$}}}
{\mbox{\boldmath $\scriptscriptstyle #1$}}}
\def\beq{\begin{equation}}
\def\eeq{\end{equation}}
\def\beqa{\begin{eqnarray}}
\def\eeqa{\end{eqnarray}}

\def\eminus{e^{-}}
\def\eplus{e^{+}}
\def\anu{\bar{\nu}}
\def\nue{\nu_e}
\def\anue{\bar{\nu}_e}
\def\numu{\nu_\mu}
\def\anumu{\bar{\nu}_\mu}
\def\nutau{\nu_\tau}
\def\anutau{\bar{\nu}_\tau}
\def\nui{\nu_i}
\def\nux{\nu_x}

\def\Enu{E_\nu}
\def\Estar{E_*}
\def\Ee{E_e}
\def\En{E_n}
\def\Enui{E_{\nui}}
\def\Enuiav{\langle\Enui\rangle}
\def\Enuisqav{\langle E^2_{\nui}\rangle}
\def\alphanui{\alpha_{\nui}}
\def\CEnuNS{CE$\nu$NS}

\def\tauplus{\tau_+}

\def\gammazero{\gamma_0}

\def\dstyle{\displaystyle}
\def\half{\frac{1}{2}}

\def\cm{\,{\rm cm}}

\def\sec{\,{\rm sec}}

\def\msun{\,{\rm M}_\odot}

\def\Fe56{^{56}{\rm Fe}}
\def\Co56{^{56}{\rm Co}}
\def\Mn56{^{56}{\rm Mn}}
\def\Pb208{^{208}{\rm Pb}}
\def\Tl208{^{208}{\rm Tl}}
\def\Bi208{^{208}{\rm Bi}} 
\def\Xe132{^{132}{\rm Xe}} 
\def\Cs132{^{132}{\rm Cs}} 

\newcommand{\GT}{\rm GT}
\newcommand{\GTminus}{{\GT}_{-}}
\newcommand{\GTplus}{{\GT}_{+}}
\newcommand{\GTzero}{{\GT}_{0}}
\newcommand{\SGTminus}{S_{\GTminus}}

\newcommand{\SGTzero}{S_{\GTzero}}
\newcommand{\F}{\rm F}
\newcommand{\SF}{S_{\F}}

\newcommand{\CC}{\rm CC}
\newcommand{\NC}{\rm NC}
\newcommand{\X}{{\rm X}}
\newcommand{\Y}{{\rm Y}}

\def\sigmanueCC{\sigma_{\nue}^{\CC}}
\def\sigmanuiNC{\sigma_{\nui}^{\NC}}
\begin{document}
\title{Supernova neutrino detection through neutron emission by nuclei}
\author{Pijushpani Bhattacharjee\inst{1} 
\fnmsep\thanks{\email{pijush.bhattacharjee@saha.ac.in}} \and 
Kamales Kar\inst{2} \fnmsep\thanks{\email{kamales.kar@gm.rkmvu.ac.in}}}
\institute{Saha Institute of Nuclear Physics, 1/AF Bidhannagar, Kolkata 
700064, India \and Ramakrishna Mission Vivekananda Educational and 
Research Institute,\\ Belur Math, Howrah 711202, India }
\abstract{Neutrinos from core collapse supernovae 
can excite nuclei of some detector materials beyond their neutron 
emission thresholds. Detection of these neutrons can give valuable 
information about the supernova explosion mechanism and 
possibly also throw light on neutrino properties. In this article, we 
give a brief review of   
the basic physics of neutrino induced neutron emission and describe the 
results of some recent calculations of supernova neutrino induced 
neutrons for some specific target detector materials due to charged 
current (CC) interactions of the electron flavored neutrinos and 
antineutrinos as well as due to neutral current (NC) interactions of 
neutrinos and antineutrinos of all flavors with the detector nuclei. We 
highlight the fact that a detector material such as lead with a 
relatively large neutron excess produces neutrons dominantly through the 
CC interaction of the $\nue$s, whereas a material such as iron with 
small neutron excess produces neutrons dominantly through the combined 
NC interaction of all the six neutrino and antineutrino
species. This raises the interesting
possibility of probing the fraction of mu- and tau flavored
neutrinos (which interact only through NC interaction) in the supernova 
neutrino flux by means of simultaneous detection of
a supernova in a lead and an iron detector, for example.} 
%
\maketitle
\section{Introduction}
\label{sec:intro}
The explosion of core collapse supernovae (SN) is associated with the 
emission of a huge number of 
neutrinos~\cite{Bethe-90,Janka-07,Scholberg:2012id}. These neutrinos 
come out from the central very high density region of 
the core of the star and are the only messengers of the conditions there
~\cite{Raffelt:1999tx,duan-09}. The detection of neutrinos from SN 1987A 
has provided direct observational evidence for neutrino emission from 
supernovae. But the number of neutrinos observed was 
very small --- a total of 20 events over a period of slightly less 
than 13 seconds at the 
two water Cerenkov facilities, Kamiokanda-II~\cite{Hirata:1987hu} and 
Irvine-Michigan-Brookhaven (IMB)~\cite{Bionta:1987qt}. However, today 
there exist many excellent neutrino detectors, like the Cerenkov or 
liquid scintillator facilities such as Super-Kamiokande~\cite{SK-11}, 
IceCube~\cite{IceCube-14}, Borexino~\cite{Cadonati-02}, 
KamLAND~\cite{Berger-01,Asakura-16}, LVD~\cite{LVD-15}, 
RENO-50~\cite{Kim-15}, and future ones like 
IceCube-Gen2~\cite{Aartsen-14}, Hyper-Kamiokande~\cite{HK-11} and  
JUNO~\cite{JUNO-16}, which will be able to detect copious number of 
neutrinos from a future nearby supernova event. While these 
observe the electron antineutrinos, the liquid argon detector within the 
DUNE facility~\cite{DUNE-15} will observe the electron neutrinos.

In this article, we discuss another mechanism of SN neutrino detection, 
namely through detection of neutrino induced neutrons with suitably 
chosen 
detector materials. Supernova neutrinos and antineutrinos interacting 
with some nuclei can excite the nuclei beyond their neutron 
emission thresholds, resulting in emission of neutrons. Neutrinos and
antineutrinos of all flavors participate in the neutral current (NC) 
excitation process, whereas the electron flavor neutrinos can 
additionally excite the nuclei through the charged current (CC) 
interaction. These neutrons detected in coincidence with the SN event 
can give valuable information about the energy spectra of 
the supernova neutrinos. Detectors for this purpose can in 
principle employ a variety of detector materials such as 
lead~\cite{Engel-03} (as in the HALO~\cite{HALO} experiment at SNOLAB),  
iron~\cite{Kolbe-01,Bandyopadhyay-17}, liquid 
xenon~\cite{Bandyopadhyay-20}, and so on. The total number of neutrons 
emitted due to CC interactions of $\nue$s with a nucleus 
\ce{^{$A$}_{$Z$}X} of mass number $A$ with $Z$ protons and $N (=A-Z)$ 
neutrons roughly scales with the ``neutron excess" ($=N-Z$) of the 
nucleus. Thus, \ce{^{208}_{82}Pb} or \ce{^{132}_{54}Xe}, for 
example, with `large' and `moderate' neutron excesses of 44 and 24, 
respectively, will respond more to electron type neutrinos than the mu 
or tau type. However, materials with low neutron excess ($N\approx Z$), 
such as $\Fe56$ (with $N=30$, $Z=26$), for example, can also be useful 
since such materials can be relatively more sensitive to $\numu$ and 
$\nutau$ and their antineutrinos (all of which only have NC interactions 
with the nuclei) because of the suppression of CC interactions of 
electron type neutrinos. As we shall see below, the total number of 
neutrons emitted also differs for the two different 
neutrino mass hierarchies, namely, Normal Ordering (NO) and Inverted 
Ordering (IO).  

In section \ref{sec:ccSN-nu}, we briefly review the general features 
of the neutrinos emitted during core collapse supernovae. Section 
\ref{sec:n-from-exc-nuc} discusses the excitation of nuclei by neutrinos 
carrying energies in the range of a few to few tens of 
MeV, both for CC and NC interactions, and the process of neutron 
emission. In section \ref{sec:n-from-diff-detector-materials} we discuss 
three specific detector materials, namely, $\Fe56$, $\Xe132$ and 
$\Pb208$ (covering the range of `low' through 
`intermediate' to `heavy' mass nuclei) with regard to their 
effectiveness for neutrino induced neutron emission, and  
present some results for these materials in section \ref{sec:results}. 
Finally, in section \ref{sec:summary}, we summarize and mention 
possible future work in this area.

\section{Neutrinos from Core Collapse Supernovae}
\label{sec:ccSN-nu}
Detailed numerical simulations with realistic physics allow one to model 
the fluxes and energies of neutrinos emitted during the 
core collapse supernovae. The emission of neutrinos can be separated 
into three distinct phases. Electron capture on nuclei $^A\X_N$ 
and on the free protons ($p$) result in
neutrino production when the shock wave responsible for the explosion 
passes through the iron core, dissociating the iron nuclei while moving 
outward. This phase, known as the
neutronization phase, lasts for 25-30 milliseconds and gives out 
predominantly electron type neutrinos ($\nue$) : 
\beq
^A\X_N + \eminus \rightarrow \, ^A\Y_{N+1} + \nue\,,
\eeq

\beq
p + \eminus \rightarrow \, n + \nue\,. 
\eeq

\begin{figure}
\epsfig{file=lum-avEnergy-timeprofile.eps,width=0.47\hsize}\hskip
0.5cm
\epsfig{file=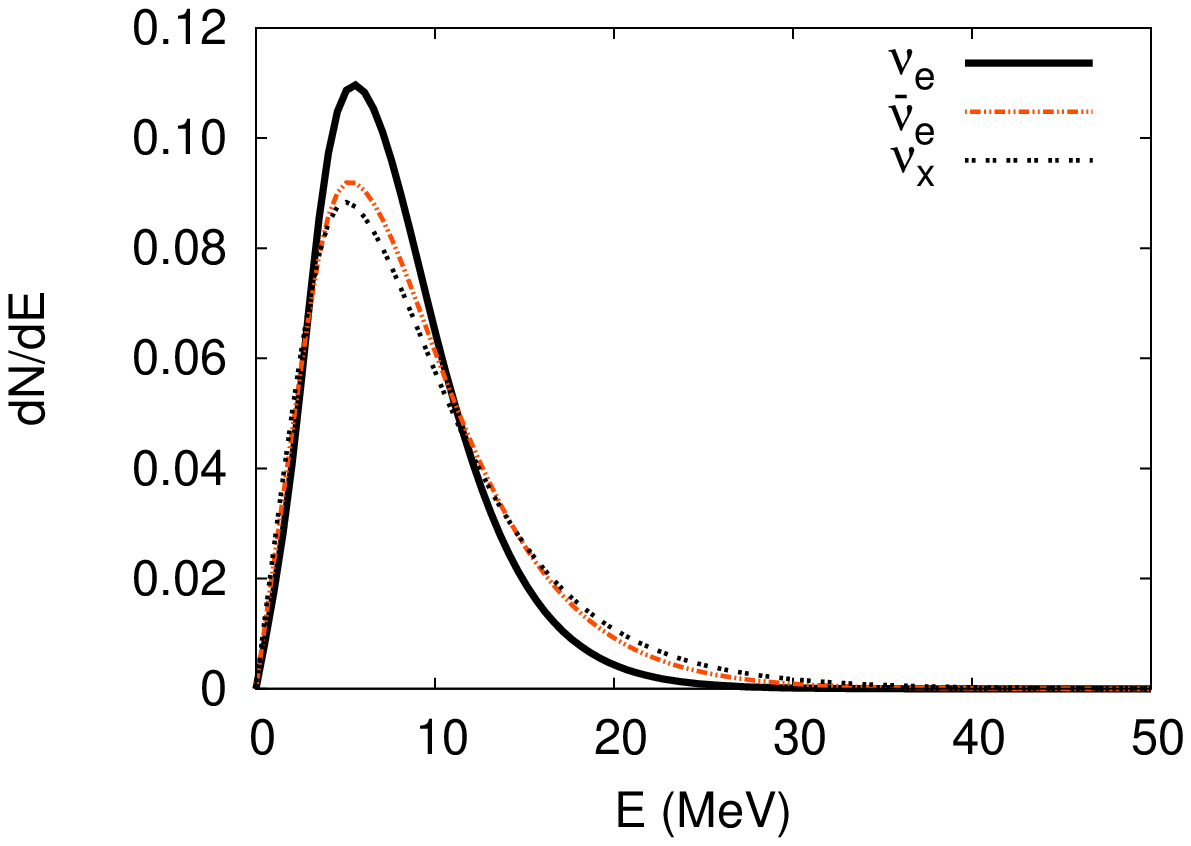,width=0.49\hsize}
\caption{Left: Temporal profiles of the neutrino luminosity
(upper three panels) and average energy of the neutrinos (lower three
panels) during the neutronization-, accretion- and
cooling phase (from left to right, respectively) for different neutrino
flavors, as given by the Basel/Darmstadt
simulation~\cite{Fischer-10} of a $18\msun$ progenitor supernova. Here 
$\nux$ represents $\numu\,$, $\anumu\,$, $\nutau\,$ or 
$\anutau\,$. Right: The normalized time-averaged energy spectra of the 
neutrinos of different flavors. (From Ref.~\cite{Bandyopadhyay-17}).}
\label{fig:lum-avEnergy-timeprofile-spectra}
\end{figure}

After this, in the accretion phase, which lasts for a few hundred 
milliseconds, neutrinos of all flavors are emitted. As the infalling 
material accretes onto the core, the matter gets sufficiently heated so 
that the $e^{+}e^{-}$ annihilation 
process results in $\nu \anu$ production of all three neutrino flavors.
The nucleon-nucleon bremsstrahlung, $ NN^{'} \rightarrow NN{'} + \nu 
\anu$, also adds to that. The $\nue$ and $\anue$ are produced through 
both CC and NC processes
whereas the $\numu$, $\anumu$, $\nutau$ and $\anutau$ are produced only
through NC processes. The $\numu$, $\anumu$, $\nutau$ and $\anutau$ 
have identical flux and energy distributions.
At the final stage neutrinos come out during the 
cooling phase which lasts for about 10 seconds during which the fluxes 
go down with time. The fluxes and average energies of $\nue$, $\anue$ 
and $\nux$ (where $\nux$ represents $\numu\,$, $\anumu\,$, $\nutau\,$ 
or $\anutau\,$) as functions of time for all the three stages from the 
results of a realistic simulation by the Basel/Darmstadt (B/D) 
group~\cite{Fischer-10} are shown in Figure 
\ref{fig:lum-avEnergy-timeprofile-spectra} for illustration. 

The average neutrino energies obtained in the B/D 
simulations~\cite{Fischer-10} are somewhat lower than those from earlier 
simulations~\cite{Totani-98,Keil-03}. In the last few 
years there has been progress in the inclusion of mean-field effects 
modifying the nuclear symmetry energy for the CC 
interaction~\cite{Martinez-12}, which goes towards reducing the 
luminosities of all neutrino flavors but increases the difference 
between the $\nue$'s and $\anue$'s average energies. There is 
also better understanding of the medium modification for neutrino-pair 
processes from nucleon-nucleon
bremsstrahlung, which modifies the process of deleptonization of the
protoneutron star~\cite{Fischer-16}. In this article we shall use for 
definiteness the temporal profiles of the neutrino luminosities 
and average energies of the neutrinos of different flavors as well as 
their normalized time-averaged energy spectra given by the 
Basel/Darmstadt simulation~\cite{Fischer-10} of a $18\msun$ progenitor 
supernova (as shown in Figure 
\ref{fig:lum-avEnergy-timeprofile-spectra}) at a distance of 10 kpc 
from earth, for illustrating our results for neutron emission in 
different detector materials.   

The post-bounce differential flux (per unit time per unit energy) of  
each neutrino flavor $\nui$ ($\nui\equiv \nue\,, \anue\,, \nux$) at time 
$t$ is written as 
\beq
F^0_{\nui} (t,\Enu) = \frac{L_{\nui} (t)}{\Enuiav
(t)}\varphi_{\nui}(\Enu,t)\,,
\label{eq:Fnuzero}
\eeq
where $L_{\nui} (t)$ and $\Enuiav (t)$ are the time dependent
luminosity and average energy of the emitted neutrinos of flavor $\nui$,
and $\varphi_{\nui}(\Enu,t)$ is the instantaneous normalized energy
spectrum ($\int \varphi_{\nui}(\Enu,t)d\Enu = 1$), which can be
parametrized as~\cite{Keil-03}
\beqa
\dstyle{\varphi_{\nui}(\Enu,t)} & = \dstyle{\frac{1}{\Enuiav (t)}
\frac{\Big(1+\alphanui (t)\Big)^{1+\alphanui (t)}}{\Gamma
\Big(1+\alphanui (t)\Big)}
\left(\frac{\Enu}{\Enuiav\left(t\right)}\right)^{\alphanui(t)}\,}
\nonumber\\
 & \,\,\,\, \dstyle{\times\,\,
\exp\left[-\Big(1+\alphanui (t)\Big)\frac{\Enu}{\Enuiav(t)}\right]\,,}
\label{eq:phi}
\eeqa
where
\beqa
\dstyle{\alphanui (t)=\frac{2\Enuiav^2(t) - \Enuisqav(t)}{\Enuisqav (t) 
- \Enuiav^2 (t)}}\nonumber
\eeqa
is the spectral shape parameter. 

It is now well-established that at least two of the three active 
neutrino species have small but non-zero masses and the probability of
finding a neutrino with a specific flavor oscillates as it 
moves~\cite{Mohapatra-Pal-book-2004}. 
These vacuum oscillation probabilities depend on the mass squared
differences and the flavor mixing angles. With the mass eigenstates
having masses $m_1$, $m_2$ and $m_3$, solar neutrino deficit
measurements give estimates of $\Delta m_{12}^2\, (\equiv m_2^2-m_1^2) 
> 0$ while the resolution of the atmospheric neutrino anomaly gives
information on the value of $|\Delta m_{23}^2|$ 
(with $\Delta m_{23}^2 \equiv m_3^2-m_2^2$), but not its sign. So one 
unsolved problem of neutrino physics is whether $m_3 > m_2$ (Normal
Ordering (NO)), or $m_3 < m_2$ (Inverse Ordering (IO)).

The vacuum oscillations get enhanced in presence of matter, an effect  
known as the Mikheyev-Smirnov-Wolfenstein (MSW) 
effect~\cite{Mohapatra-Pal-book-2004}. For neutrinos
moving through matter with a density gradient a resonance in the
oscillations can take place. The neutrinos can have the resonance
for NO but not for IO and similarly the antineutrinos can have the
resonance for IO, but not for NO. 

For core collapse supernovae the matter density goes through a large
gradient from the central region to the edge of the core. Generally
one encounters two densities where resonances can take place
\cite{Dighe-2000} and so the luminosities of the different neutrino
flavors depend on the mass ordering (NO or IO). So the flux of neutrinos
reaching the detector is different for different mass ordering. In 
addition, of course, the flux of neutrinos reaching the earth has the 
factor of $1/(4 \pi d^2)$ where $d$ is the distance of the 
supernova from the earth.

On top of the matter-enhanced oscillations, there can be oscillations 
due to neutrino-neutrino interaction or neutrino 
self-interaction, an effect known as collective 
oscillations \cite{Duan-10,Mirrizi-16}. This happens only at
regions with very high neutrino densities in the region of a few hundred
kms from centre of the core during the accretion phase at a post-bounce 
time $t_{\rm pb} < 0.5\,\sec$. This results in multiple splits in 
the neutrino spectra. This pre-processes the fluxes of the neutrinos 
before they move into the MSW region. However as the net electron 
density in this collective oscillation region is not small, 
one sees that large matter induced phase 
dispersion for neutrinos traveling in different 
directions partially or totally suppress the collective 
oscillations~\cite{Saviano-12}. Detailed numerical 
simulations~\cite{Chakraborty-11a,Chakraborty-11b} observed this 
suppression using results of SN hydrodynamic simulations. On the 
other hand, in the cooling phase, the matter suppression is not present 
anymore and collective effects should take place before the 
neutrinos enter the regions of lower densities. But in the cooling phase 
the neutrino fluxes of different flavors are very similar
and the effects of such oscillations on them are not important from the 
observational point of view. So, for simplicity, in this article we 
ignore the collective oscillation effect and consider only the MSW 
oscillation effects. 

\section{Nuclear Excitations By Neutrinos and Emission of Neutrons}
\label{sec:n-from-exc-nuc}
Charged current interaction of a SN $\nue$ with a detector nucleus 
$^A_Z\X_N$ can produce the nucleus $^A_{Z+1}\Y_{N-1}$ which, 
depending on the incident $\nue$ energy, can be produced in an  
excited state (denoted by a superscript $*$): 
\beq
\nue + _{Z}^{A}\X_{N} \rightarrow \, \eminus + \, 
^A_{Z+1}\Y^{*}_{N-1}\,,
\label{eq:nue-CC-general} 
\eeq
with subsequent de-excitation of the final state nucleus through 
emission of various particles ($\gamma$, $p$, $n$, $\alpha$, 
and so on) depending on the excitation energy of the nucleus. We are 
interested in the situation when the nucleus de-excites by 
emitting one or more neutrons: 
\begin{eqnarray}
^A_{Z+1}\Y^{*}_{N-1} & \to & ^{A-1}_{Z+1}{\Y}_{N-2} + n\,,
\label{eq:nue-CC-general-1n-emission}\\
^A_{Z+1}\Y^{*}_{N-1} & \to & ^{A-2}_{Z+1}{\Y}_{N-3} + 2 n\,,
\label{eq:nue-CC-general-2n-emission}
\end{eqnarray}
and so forth. Emission of three or more neutrons is possible in 
principle but their contribution 
to the total number of neutron emission is negligibly small as
the thresholds for emission of three or more neutrons are generally very 
high compared to the range of energies over which the nucleus can be 
excited by the SN neutrinos.   

Similar to $\nue$s, the SN $\anue$s can also interact with the 
detector nucleus through CC interaction producing a $\eplus$ 
and an excited final state nucleus: 

\begin{eqnarray}
\anue + ^A_Z{\X}_N \to\, \eplus + ^{A}_{Z-1}{\tilde\Y}^{*}_{N+1}\,.
\label{eq:anue-CC-general}
\end{eqnarray}
The final state nucleus, if excited above single, double or higher 
neutron emission thresholds will emit neutrons in competition with other
particles. However, in this case, since the production 
of the final state nucleus involves conversion of a proton into a 
neutron inside the nucleus, the cross section for the process is 
strongly suppressed due to Pauli blocking of the neutron single particle
states in nuclei with moderate to large neutron excesses ($N-Z$) that we 
shall consider. Contribution of $\anue$'s to neutron emission is 
therefore much smaller than that due to $\nue$s.   

Neutrinos and antineutrinos of all three flavors ($\nui$) can also 
excite the target nucleus through the flavor blind NC process whereby 
the incoming $\nu$ or $\anu$ inelastically scatters off the nucleus 
leaving the latter in an excited state,  
\beq
\nui + ^{A}_{Z}\X_{N} \rightarrow\, \nui + ^{A}_{Z}\X^{*}_{N}\,,
\label{eq:nui-NC-general}
\eeq
with subsequent de-excitation of the final state nucleus through neutron 
emission. Thus NC process has contribution from all six types of 
neutrinos whereas CC has contribution essentially from only one.

\subsection{Neutrino-Nucleus charged current cross section}
\label{subsec:CC-xsec-general}
For the range of energies of supernova neutrinos, the $\nue$ CC 
cross section in the $q \rightarrow 0$ limit ($q$ being the momentum 
transfer) is dominated by the two allowed transitions, (a) the Fermi 
transition (given by $\sum_{i}\tauplus(i)$, where $\tauplus$ is the 
operator that converts a neutron to a proton and the summation is over 
all the nucleons), which 
goes almost completely to the Isobaric Analog State (IAS) of the final 
nucleus, and (b) the Gamow-Teller (GT) transition (given by the 
operator $\sum_{i}{\bg\sigma}(i)\tauplus(i)$ where $\bg\sigma(i)$
are the standard Pauli spin matrices representing the spin operator for 
the $i$-th nucleon), which is spread over a broad resonance over many 
final states with overlapping strengths, with some small part going to a 
few discrete low-lying states. 

Thus the $\nue$ CC differential cross section in the $q\to 0$ limit 
can be written 
as~\cite{Kuramoto-etal-90,Fuller-etal-99,Kolbe-etal-99}
\beq
\frac{d\sigmanueCC}{d\Estar}(\Enu,\Estar) =
\frac{G_F^2 \cos^2\theta_c}{\pi} \,\,\, p_e E_e F(Z+1,E_e)\,
  \left[\SF(\Estar) + 
(g^{\rm eff}_{\rm A})^2 \SGTminus(\Estar)\right]\,,
\label{eq:nue_CC_diff_xsec}
\eeq
where $G_F$ is the Fermi constant, $\theta_c$ is the Cabibbo angle,
$p_{e}$ and $E_e$ are the momentum and energy of the emitted electron,  
$\Estar=\Enu-\Ee$ is the excitation energy of the final nucleus, and 
$\SF(\Estar)$ and $\SGTminus (\Estar)$ are respectively 
the averaged Fermi and Gamow-Teller ($\GTminus$) matrix elements between 
the ground state of the initial nucleus $^A_Z\X_N$ and the excited state 
(at energy $\Estar$) of the final nucleus $^A_{Z+1}\Y^{*}_{N-1}$:  
\beq
\SF(\Estar)=\frac{1}{2J_i+1}\big\vert \langle
J_f\|\sum_{k=1}^{A}\tauplus(k)\|J_i\rangle\big\vert^2 
\label{eq:S_F}
\eeq
and
\beq
\SGTminus(\Estar)=\frac{1}{2J_i+1}\big\vert \langle
J_f\|\sum_{k=1}^{A}\tauplus(k) {\bg\sigma}
(k)\|J_i\rangle\big\vert^2\,.
\label{eq:S_GT}
\eeq
The quantity $g^{\rm eff}_{\rm A}\simeq 1.26$ is the ratio of the
effective axial vector to vector coupling constants of the bare 
nucleon in the $q\to 0$ limit~\cite{Kuramoto-etal-90,Fuller-etal-99}. 
In equation (\ref{eq:nue_CC_diff_xsec}) the factor $F(Z+1,E_e)$, which 
takes into account the Coulomb distortion 
of the outgoing electron wave function, is given by~\cite{Engel-98} 
\beq
F(Z,E)=2 (1+\gammazero)(2p_e R)^{2(\gammazero-1)}
\frac{\arrowvert\Gamma(\gammazero+iy)\arrowvert^2}{
\arrowvert\Gamma(2\gammazero+1)\arrowvert^2}\exp(\pi y)\,,
\label{eq:FermiFunction}
\eeq
where $\gammazero=(1-Z^2\alpha^2)^{1/2}$, $y=\alpha Z E_e/p_e$, $R$ 
is the radius of final nucleus and $\alpha$ the fine structure constant.

There are also contributions from forbidden transitions ($\Delta l \neq 
0$ for the single particle transitions), but normally they are one to 
two orders of magnitude smaller than 
the contributions of the allowed ones and need to be considered 
only when the Fermi and most of GT strengths are blocked.

Phenomenologically it is seen that the bare nucleon value for 
$g^{\rm eff}_{\rm A}\simeq 1.26$ somewhat overestimates the GT strengths 
at low excitation energies. But since for our purpose we are interested 
in the GT strengths at higher excitations beyond the neutron emission 
thresholds covering the whole GT resonance over tens of MeV, we think 
the use of the bare nucleon value for $g^{\rm eff}_{\rm A}$ is a good 
approximation. 

The total $\nue$ CC cross section as a function of the 
incoming neutrino energy $\Enu$ is then given by
\beq
\sigmanueCC (\Enu) = \int_0^{\Enu}\frac{d\sigmanueCC}{d\Estar} 
d\Estar\,.
\label{eq:sigma_nue_CC}
\eeq

For the GT strength distribution one often does theoretical calculations 
in model many-nucleon spaces with realistic interactions which can 
reproduce the observed ground state and excited state energies as well 
as the observed $\log ft$ values for a few low-lying states. For some 
nuclei forward angle $(p,n)$ reaction gives the $\GTminus$ strengths 
experimentally. For comparatively lighter nuclei in the Fe-Ni region, 
one does a large dimensional shell model calculation \cite{Caurier-99} 
or a calculation using the Shell Model Monte Carlo (SMMC) technique 
\cite{Langanke-95}. The strengths often sensitively depend on the 
interaction. For the lower part of $fp$-shell nuclei, the 
Kuo-Brown KB3~\cite{Kuo-Brown-68,Poves-81} 
interaction with the corrected monopole interaction is adequate. 
However the upper part of the shell needs a better handling of the 
monopole part involving the higher orbits in the shell. but this 
shell model approach becomes computationally increasingly difficult as 
one goes to heavier nuclei. For these nuclei, like $\Pb208$, one 
resorts to the RPA approach~\cite{Engel-03}. Finally, for nuclei which 
are not spherical in ground state region one carries out a deformed 
Skyrme Hartree-Fock mean field calculation with pairing correlation in 
the BCS approximation~\cite{Sarriguren-98}. The $\GTminus$ transition 
strengths are obtained by a quasiparticle random phase approximation 
(QRPA) with a residual spin-isospin interaction~\cite{Moreno-06}.

\subsection{Neutrino-Nucleus neutral current cross section}
\label{subsec:NC-xsec-general}
For the NC, the allowed contribution to the differential cross section 
for inelastic scattering involving energy transfer to the target nucleus 
in the $q\to 0$ limit is again governed by the corresponding GT (the 
so-called $\GTzero$) strength~\cite{Fuller-etal-99}, and is given by 
\beq
\frac{d\sigmanuiNC}{d\Estar}(\Enu,\Estar) = E_{\nu^{\prime}}^2\,
(g^{\rm eff}_{\rm A})^2 \,\SGTzero(\Estar)\,,
\label{eq:nui_NC_diff_xsec}
\eeq
where $E_{\nu^{\prime}}=\Enu - \Estar$ is the energy of the final 
neutrino. The GT strength $\SGTzero$ involves the `$z$'-component of the 
isospin vector of the GT operator, and is given by
\beq
\SGTzero(\Estar)=\frac{1}{2J_i+1}\big\vert \langle
J_f\|\sum_{k=1}^{A}\half\tau_3 (k) {\bg\sigma}
(k)\|J_i\rangle\big\vert^2\,. 
\label{eq:S_GT0}
\eeq
The analog of Fermi contribution in this case contributes only to the 
elastic part. In some cases where energy 
considerations block most of the allowed strength, contributions from 
the much smaller forbidden transitions need to be taken into account.

The total $\nui$ NC cross section is obtained by integrating the 
differential NC cross section (\ref{eq:nui_NC_diff_xsec}) over the 
excitation energy $\Estar$ of the final nucleus.  

\subsection{Emission of neutrons by excited nuclei}
As already implicitly assumed above, the neutrino induced neutron 
emission from nuclei through $\nue$ or $\anue$ CC interaction and $\nui$ 
NC inelastic scattering can be considered as a 2-step process, with 
the final state nucleus being produced in an excited state due to 
absorption of energy from the incoming neutrino in the first step, and 
subsequent de-excitation of the nucleus through neutron emission (if the 
excitation energy is above the threshold for neutron emission) in the 
second step. The physical processes involved in these two steps can be 
considered to be independent of each 
other~\cite{Kolbe-01,Engel-03}. The differential cross 
sections for the first step, i.e., nuclear excitation  
through neutrino-nucleus CC or NC interaction, are given by 
equations (\ref{eq:nue_CC_diff_xsec}) (for $\nue$) and 
(\ref{eq:nui_NC_diff_xsec}), respectively. The cross section for 
emission of one-, two- or three neutrons, for example, by the final 
nucleus is then given by  

\beq
\sigma_{1n (2n) (3n)}^{\rm CC (NC)} (\Enu)=\int
\frac{d\sigma^{\rm CC (NC)}}{d\Estar}(\Enu,\Estar) P_{1n (2n)
(3n)}(\Estar)d\Estar\,,
\label{eq:n-emission-xsec}
\eeq
where $\frac{d\sigma^{\rm CC (NC)}}{d\Estar} (\Enu,\Estar)$ represents 
the 
relevant differential cross section given by equation 
(\ref{eq:nue_CC_diff_xsec}) or (\ref{eq:nui_NC_diff_xsec}), and 
$P_{1n}(\Estar)$, $P_{2n}(\Estar)$, $P_{3n}(\Estar)$ are the 
probabilities for emission of one-, two- and three neutrons, 
respectively, by the final nucleus, as functions of the excitation 
energy of the final nucleus. 

One can also calculate the energy spectrum of the emitted neutrons in 
the following way: The differential cross section for emission of 
neutrons per unit neutron energy $\En$ by a single nucleus due to an 
incoming neutrino of energy $\Enu$ can be written as 
\beq
\frac{d\sigma^{\rm CC (NC)}}{d\En}(\Enu,\En)
= \int \frac{d\sigma^{\rm CC (NC)}}{d\Estar}(\Enu,\Estar) 
\frac{dN_n}{d\En}(\Estar,\En)\, d\Estar\,,
\label{eq:dsigma-n_dEn}
\eeq
where $\frac{dN_n}{d\En}(\Estar,\En)$ is the energy spectrum of the 
neutrons produced by the excited nucleus of excitation energy $\Estar$, 
with 
\beq
\int \frac{dN_n}{d\En}(\Estar,\En)\, d\En = P_{1n}(\Estar) + 
2\,P_{2n}(\Estar) + 3\, P_{3n}(\Estar)\,,
\label{eq:dNn_dEn}
\eeq
considering up to 3-neutron emission. 

The energy spectrum of all the neutrons produced by the incident 
flux of SN neutrinos is then given by 
\beq
\frac{dN_{n,{\rm total}}^{\rm CC (NC)}}{d\En}=N_0 \int d\Enu 
\Phi_{\nu}(\Enu) 
 \frac{d\sigma^{\rm CC (NC)}}{d\En}(\Enu,\En)\,,
\label{eq:total-n-spect}
\eeq
where $\Phi_{\nu}(\Enu)$ is the time-integrated flux spectrum (number 
per unit area per unit energy) of the SN neutrinos falling on the 
detector, and $N_0$ is the total number of target detector nuclei.

Finally, the total number of neutrons produced is given by 
\beq
N_{n,{\rm total}}^{\CC (\NC)} = \int d\En
\frac{dN_{n,{\rm total}}^{\rm CC(NC)}}{d\En}\,= N_0 \int d\Enu 
\Phi_{\nu}(\Enu) \sigma_{n,{\rm total}}^{\CC(\NC)} (\Enu)\,,
\label{eq:n-total-number}
\eeq
where 
\beq
\sigma_{n,{\rm total}}^{\CC(\NC)} (\Enu)\equiv 
\sigma_{1n}^{\CC(\NC)} (\Enu) + 2\sigma_{2n}^{\CC(\NC)} (\Enu) + 
3\sigma_{3n}^{\CC(\NC)} (\Enu)
\label{eq:sigma-n-total-def}
\eeq
with $\sigma_{1n (2n) (3n)}^{\rm CC (NC)} (\Enu)$ given by equation
(\ref{eq:n-emission-xsec}). 

\begin{figure}
\begin{center}
\includegraphics[scale=.5]{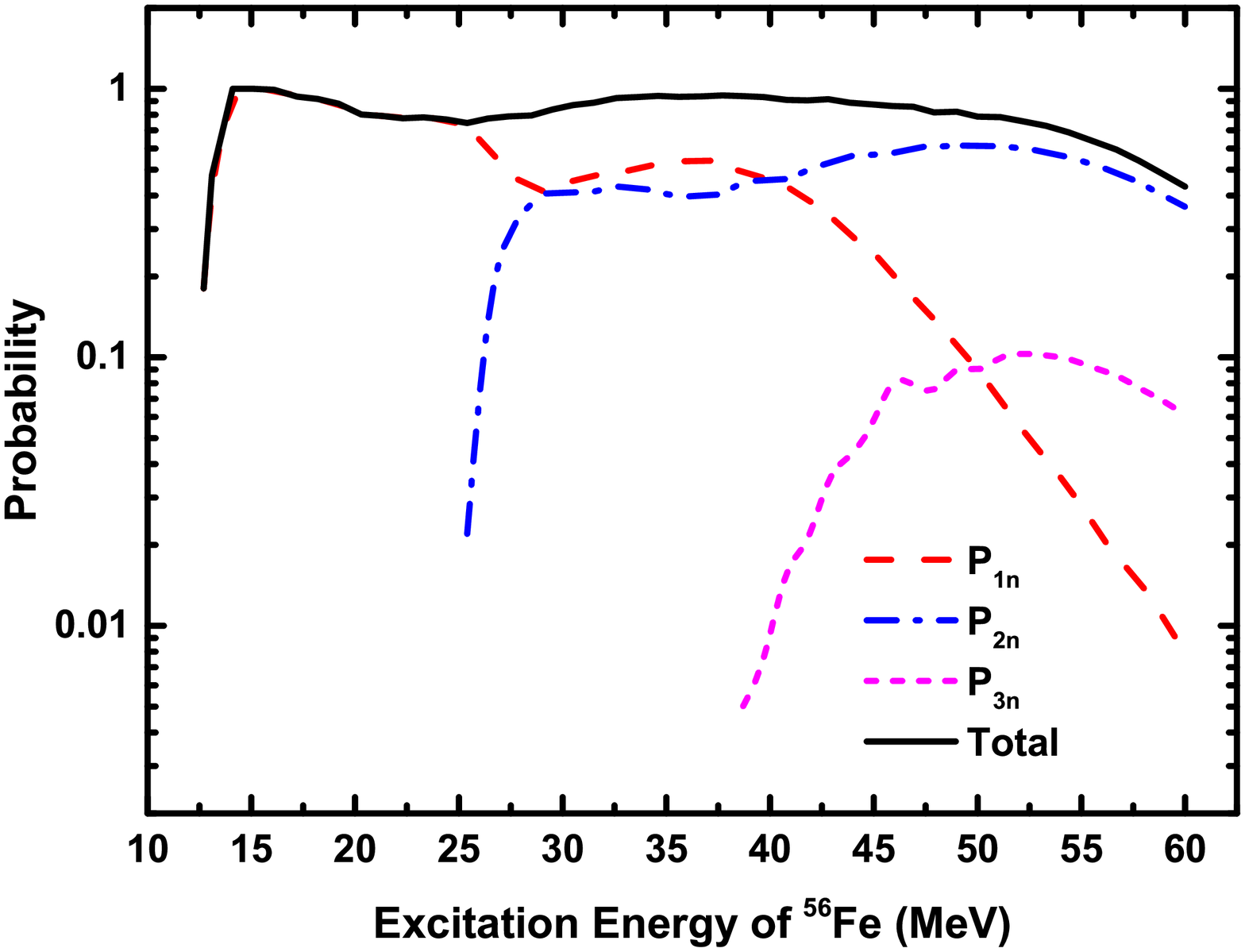}\\
\caption{The one-, two- and three neutron emission probabilities
of excited $\Fe56$ nucleus calculated with the PACE4 
code~\cite{PACE4-code}. (From ~\cite{Bandyopadhyay-17}.)}
\label{fig:n-emission-prob-exc-56Fe}
\end{center}
\end{figure}
\begin{figure}
\begin{center}
\includegraphics[scale=0.5]{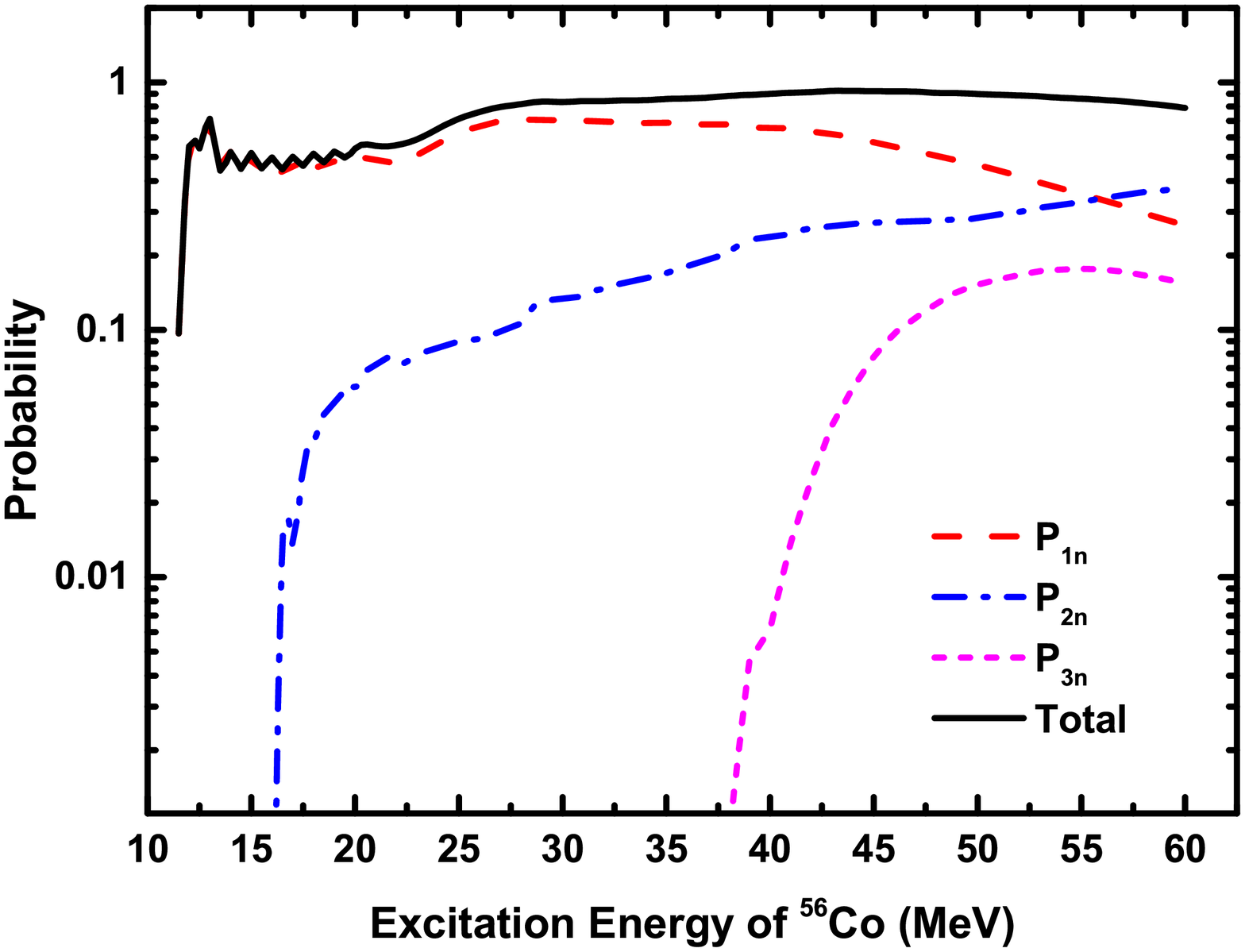}\\
\caption{The one-, two- and three neutron emission probabilities of
excited $\Co56$ nucleus calculated with the PACE4 
code~\cite{PACE4-code}. (From ~\cite{Bandyopadhyay-17}.)}
\label{fig:n-emission-prob-exc-56Co}
\end{center}
\end{figure}

The neutron emission probabilities, $P_{1n (2n) (3n)}(\Estar)$,  
and the neutron energy spectrum, $\frac{dN_n}{d\En}(\Estar,\En)$, for 
the desired excited nucleus 
can be calculated using the fusion-evaporation code 
PACE4~\cite{PACE4-code} originally developed
by Gavron~\cite{Gavron-80}. The neutron emission probabilities from
excited $\Fe56$ and $\Co56$ nuclei as functions of excitation
energy calculated with the PACE4 code are shown in 
Figures~\ref{fig:n-emission-prob-exc-56Fe} and 
\ref{fig:n-emission-prob-exc-56Co}, respectively, for illustration. 

\section{Neutron emission from different detector materials}
\label{sec:n-from-diff-detector-materials}
The underground SNOLAB facility in the Creighton Mine in Sudbury, Canada
houses a detector named HALO (Helium and Lead Observatory)~\cite{HALO} 
which is dedicated for supernova detection through detection of 
neutrino induced neutrons. The HALO detector consists of 79  
tons of lead and uses a large number of $^{3}{\rm He}$ neutron 
detectors. The element $\Pb208$ with $Z=82$ protons has a neutron excess 
($N-Z$) of 44 and thus the total $\GTminus$ strength is large as 
it roughly scales as $3(N-Z)$ 
according to Ikeda sum rule, with the $\GTplus$ very small. The 
$\GTplus$ strength coming from excitation of the nucleus by $\anue$ is
extremely small for $\Pb208$ as the protons need to move to states 
across a major shell in the shell model picture of single particle 
states, and so allowed $\GTplus$ transitions are suppressed due to 
Pauli blocking. In addition, the contributions from the order of 
magnitude smaller forbidden transitions are
also very small. For the CC reaction $\Pb208\,(\nue,\eminus)\Bi208$, 
the excited final nucleus $\Bi208$ has one and two neutron emission 
thresholds at 6.89 and 14.99 MeV, respectively. 
As the $\GTminus$ strength is a broad resonance spread over tens of 
MeVs, most of the Gamow-Teller transitions from $\Pb208$
and the allowed Fermi transition contribute to the process of neutron 
emission.

For the NC interaction $\Pb208\,(\nui,\nui)\Pb208$ the one and two 
neutron emission thresholds have comparatively low values of 7.37 and 
14.11 MeV, respectively, and in this case 
all six neutrino species contribute. Moreover, $\Pb208$ also has a low 
neutron capture cross section as it is a doubly magic nucleus. Thus the 
neutrons emitted from the excited nuclei resulting from both CC and NC 
interactions of the SN neutrinos survive while moving through the 
bulk detector material and are able to reach the neutron detectors. 

The iron detectors with a much lighter nucleus $\Fe56$ with $Z=26$ has 
the disadvantage of having a small neutron excess ($N-Z$) of only 
4, and consequently its total $\GTminus$ strength is an order 
of magnitude smaller than in the case of lead. On
top of that, for the CC reaction $\Fe56\,(\nue,\eminus)\Co56$, the 
final $\Co56$ has the one and two neutron emission thresholds quite high 
at 10.08 and 24.17 MeV, respectively. So the Fermi 
transition to the Isobaric Analog State (IAS) gets blocked and only a 
part of the $\GTminus$ transitions can contribute.

The total $\GTplus$ strength for the process 
$\Fe56\,(\anue,\eplus)\Mn56$ is smaller than the total 
$\GTminus$ strength for $\Fe56\,(\nue,\eminus)\Co56$, as the valence 
protons of $\Fe56$ have to go to 
neutron single particle states at much higher energies. The total 
$\GTplus$ strength for $\Fe56$ has the observed value of $\sim 2.8$ 
while the shell model calculations give a value of 
2.7~\cite{Caurier-99}. In contrast the total $\GTminus$ strength is 
$\sim$ 9.9 experimentally and $\sim$ 9.3 from theory \cite{Caurier-99}. 
Here again the final nucleus $\Mn56$ in the reaction  
$\Fe56\,(\anue,\eplus)\Mn56$ has one neutron emission threshold 
above 8 MeV whereas the $\GTplus$ strength distribution is spread over 
excitation energies below 8 MeV. Hence very 
little contribution comes from SN $\anue$ absorption and it can be
safely neglected. 

Coming to intermediate mass nuclei like xenon, we note that liquid xenon 
scintillator detectors are already in use for dark matter (DM) detection 
experiments~\cite{DM-detection-review-16}. 
These experiments attempt to detect nuclear recoils that would be caused 
by the Weakly Interacting Massive Particle (WIMP) candidates of DM.  
However, such liquid xenon DM detectors, because of their low nuclear 
recoil thresholds, can also respond to supernova 
neutrinos~\cite{Chakra-14,Lang-16} through the process of coherent 
elastic neutrino-nucleus scattering 
(\CEnuNS)~\cite{Freedman-74-77,Horowitz-etal-03}. At the same time CC 
and NC interactions of SN neutrinos with xenon nuclei can also result in 
excitation of the final state nucleus, the subsequent de-excitation of 
which can result in emission of neutrons which can be 
detected through xenon nuclear recoils caused 
by these neutrons~\cite{Bandyopadhyay-20}. 

Quite a few isotopes of xenon are stable with even mass isotopes like
$^{124}{\rm Xe}$, $^{126}{\rm Xe}$, $^{128}{\rm Xe}$, $^{130}{\rm Xe}$, 
$\Xe132$, $^{134}{\rm Xe}$ and $^{136}{\rm Xe}$ having abundances 
0.095\%, 0.089\%, 1.910\%, 4.071\%, 26.909\%, 10.436\% and 8.857\%,  
respectively. Among the odd mass ones $^{129}{\rm Xe}$ and 
$^{131}{\rm Xe}$ have abundances of 26.40\% and 21.232\% respectively. 
In this article we consider the example of $\Xe132$ and find its 
effectiveness for neutron emission after excitation by supernova 
neutrinos. The $\Xe132$ with $Z=54$ has a neutron excess ($N-Z$) of 24, 
and with the total $\GTplus$ strength for the reaction 
$\Xe132\,(\anue,\eplus)^{132}{\rm I}$ being very small due to 
Pauli blocking of single particle transitions, the $\GTminus$ total 
strength for the $\nue$ CC reaction $\Xe132\,(\nue,\eminus)\Cs132$ is 
expected to be $\gsim72 \,(=3(N-Z))$ by the Ikeda sum rule. 
This is indeed predicted by theoretical calculations~\cite{Moreno-06}
where total $\GTminus$ is obtained as 72.12 and total $\GTplus$ as 0.51.
Also the Fermi strength of 24.0 goes to the IAS in $\Cs132$ at an 
excitation of $\sim$ 13.8 MeV~\cite{Unlu-16}. Finally, the one and two 
neutron emission thresholds for the nucleus $\Cs132$ are at 7.17 and 
16.40 MeV, respectively. Thus in the CC 
reaction on $\Xe132$ the Fermi transition and most of the 
$\GTminus$ transition strength contribute like in the case of $\Pb208$.

\section{Results}
\label{sec:results}
\begin{figure}
\begin{center}
\includegraphics[scale=.5]{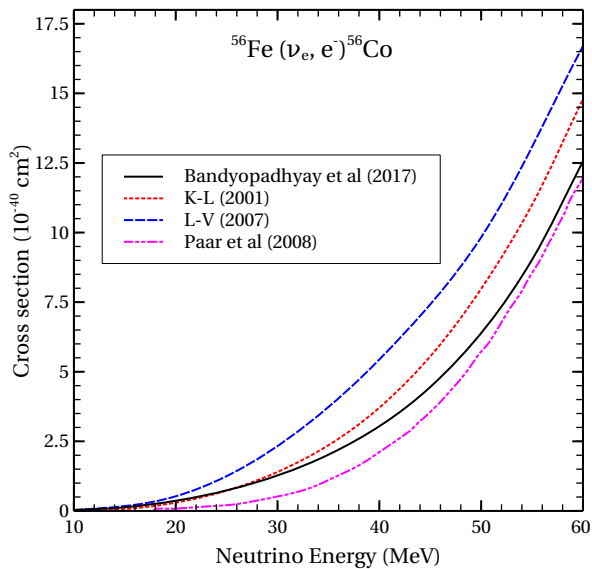}\\
\caption{Comparison of the CC cross section for the process 
$\Fe56\,(\nue,\eminus)\Co56$ calculated in four different works, 
Bandyopadhyay et al~(2017)~\cite{Bandyopadhyay-17}, 
Kolbe and Langanke~\cite{Kolbe-01} (K-L (2001)), Lazauskas and
Volpe~\cite{Laz-07} (L-V (2007)), and Paar et al~\cite{Paar-08} (Paar et 
al (2008)). (From Ref.~\cite{Bandyopadhyay-17}).}
\label{fig:fexseccompare-56Fe}
\end{center}
\end{figure}

The $\nue$ CC cross section on $\Fe56$ has been calculated 
using the $\GTminus$ energy distribution generated by different 
theoretical calculations. In Figure \ref{fig:fexseccompare-56Fe} we show 
a comparison of four such results where the cross section as a function 
of the neutrino energy is shown. 

The CC cross section for $\Fe56\,(\nue,\eminus)\Co56$ calculated by 
Bandyopadhyay et al~\cite{Bandyopadhyay-17} uses the results of large
dimensional shell model calculations of the $\GTminus$ strength from the 
$0^{+}$ ground state of $\Fe56$ to all $1^{+}$ states of 
$\Co56$~\cite{Caurier-99}. The calculations of Ref.~\cite{Caurier-99} 
use the very successful monopole-corrected KB3 
interaction~\cite{Kuo-Brown-68,Poves-81}. In 
addition, Bandyopadhyay et al ~\cite{Bandyopadhyay-17} include the 
dominant contributions coming from the forbidden
transitions to the $1^{-}$ and $2^{-}$ states of $\Co56$. The $\nue$ CC
cross section of this calculation agrees well with the results
of Kolbe and Langanke (K-L)~\cite{Kolbe-01} particularly at lower  
energies. The K-L work for the transition strengths to 
$1^{+}$ states of $\Co56$ also use an interacting shell model 
calculation within the complete $pf$ shell with an overall
quenching factor of $(0.74)^{2}$ for the strength distribution.  
The results of Lazauskas and Volpe (L-V)~\cite{Laz-07} give the largest
cross sections for all neutrino energies. They carry out a Hartree-Fock (HF)
calculation from the $\Fe56$ ground state to the occupied single 
particle states using the Skyrme-type effective interaction. The 
unoccupied levels are obtained by diagonalizing the HF mean field using 
a harmonic oscillator basis. Also when needed pairing correlations are 
taken into account in the HF+BCS approximation. Finally in the work by 
Paar, Vretenar and Ring~\cite{Paar-08} the nuclear ground state is 
described by the relativistic Hartree-Bogoliubov model and the 
transition strengths to the excited levels in the relativistic 
quasiparticle random phase approximation (RQRPA). Their CC 
cross section values are lower than the ones in the other three
calculations for the entire range of neutrino energy. 

In Table \ref{tab:nue-CC-neutron-xsec} we 
compare the neutron emission 
cross sections for $\Pb208$, $\Xe132$ and $\Fe56$ for neutrino energies 
in the range of SN neutrino spectra. The values for all three cases 
increase with the neutrino energy as the differential cross section  
given by equation (\ref{eq:nue_CC_diff_xsec}) is proportional to the 
product of the electron energy and the electron momentum and their range 
of values increases as $E_{\nu}$ increases.
But we see that the cross section values for $\Fe56$ are 
significantly smaller than those for the other two nuclei. 
As discussed earlier this is due to $\Fe56$ having very 
small neutron excess and the high neutron emission threshold of $\Co56$ 
blocking the contribution from the Fermi strength and a part of the 
$\GTminus$ strength. The intermediate mass nucleus 
$\Xe132$ with a neutron excess of 24 has cross sections roughly half the 
values for $\Pb208$ which has a neutron excess of 44. Note also that for 
$\Pb208$ beyond  30 MeV the 2n contribution 
becomes comparable to 1n values. But for $\Fe56$ the 1n contribution 
dominates over the 2n values over the whole range of neutrino energy 
considered.
\begin{table}
\begin{center}
\caption{The total neutron emission cross section,  
$\sigma^{\CC}_{n,{\rm total}}$ 
(defined by eq.~(\ref{eq:sigma-n-total-def})),
in units of $10^{-40} \cm^{2}$, for different neutrino energies  
due to $\nue$ CC interaction with $\Pb208$ (from \cite{Engel-03}), 
$\Xe132$ (from \cite{Bandyopadhyay-20}) and 
$\Fe56$ (from \cite{Bandyopadhyay-17}).  
These include contributions from one- and two neutron emissions with   
cross sections respectively given within parentheses, 
$(\sigma^{\CC}_{1n}\,,\sigma^{\CC}_{2n})$, for $\Pb208$ and $\Fe56$.  
(The $1n$ and $2n$ emission cross sections values 
for $\Xe132$ will be reported in \cite{Bandyopadhyay-20}.)     
The very small contributions from three 
neutron emission is neglected, so $\sigma^{\CC}_{n,{\rm total}}\approx 
\sigma^{\CC}_{1n} + 2 \sigma^{\CC}_{2n}$. These total neutron emission 
cross sections, folded with the incident $\nue$ flux, give the total 
number of emitted neutrons for the chosen detector material, 
respectively (see eq.~(\ref{eq:n-total-number}))}. 
\label{tab:nue-CC-neutron-xsec} 
\begin{tabular}{|c|c|c|c|}
\hline 
$\Enu$ (MeV) & $\Pb208$  & $\Xe132$ & $\Fe56$  \\
\hline
\hline
5   &   0.0   &   0.0    &    0.0   \\
\hline 
10  &   0.0   &   0.04   &    0.0   \\
\hline
15  &   0.91 (0.91, 0)  &   0.13   &    0.0   \\
\hline
20  &   4.96 (4.96, 0) &   2.09   &    0.03 (0.03, 0)  \\
\hline
25  &   15.56 (14.66, 0.45) &   7.33   &    0.07 (0.07, 0) \\
\hline
30  &   31.35 (25.05, 3.15) &  15.78   &    0.15 (0.15, 0) \\
\hline
35  &   50.97 (29.27, 10.85) &  27.36   &    0.29 (0.28, 0.005) \\
\hline
40  &   80.92 (33.56, 23.68) &  41.91   &    0.53 (0.49, 0.02)   \\
\hline
45  &  115.85 (37.91, 38.97) &  59.29   &    0.93 (0.85, 0.04)  \\
\hline
50  &  150.12 (42.54, 53.79) &  79.37   &    1.58 (1.40, 0.09) \\
\hline
55  &  190.43 (47.17, 71.63) & 102.14   &    2.57 (2.23, 0.17) \\
\hline
60  &  232.12 (52.02, 90.05) & 127.78   &    4.01 (3.43, 0.29) \\
\hline
\end{tabular}
\end{center}
\end{table}

Next we come to the results for the number of neutrons emitted. As 
already mentioned, due to neutrino flavor oscillation the flux of SN 
neutrinos at earth is different for the two different neutrino mass 
orderings, NO and IO. As a result the numbers of neutrons emitted are  
also different for the two cases. Table 
\ref{tab:n-numbers_vs_Enu-bins_nue-CC_NO-IO} gives the number of 
emitted neutrons for three different detector materials, namely, 
$\Fe56$, $\Xe132$ and $\Pb208$, for CC interaction of the 
SN $\nue$s, for both NO and IO mass orderings. The numbers are for SN 
neutrino flux as given by the Basel/Darmstadt 
simulations~\cite{Fischer-10} of a $18\msun$ progenitor supernova at a 
distance of 10 kpc, and for 1 kton of the given detector material.  
These numbers are obtained by folding the neutron 
emission cross sections given in Table \ref{tab:nue-CC-neutron-xsec} 
with the corresponding SN $\nue$ flux at earth. 

\begin{table}
\begin{center}
\caption{Number of neutrons $(N_n)$ emitted per kton of different 
detector materials along with the cumulative total number (Sum)  
as a function of $\nue$ energy (in 5 MeV bins) due to CC 
interactions of the SN-$\nue$s, 
for the case of Normal Ordering (NO) of neutrino masses, with the 
corresponding numbers for the Inverse Ordering (IO) given in 
parentheses. The numbers are for SN neutrino flux as given by the 
Basel/Darmstadt simulations~\cite{Fischer-10} of a $18\msun$ 
progenitor supernova at a distance of 10 kpc. The numbers for 
$\Fe56$ and $\Pb208$ are from Ref.~\cite{Bandyopadhyay-17} and those 
for $\Xe132$ are from Ref.~\cite{Bandyopadhyay-20}.}    
\label{tab:n-numbers_vs_Enu-bins_nue-CC_NO-IO} 
\begin{tabular}{|c|c|c||c|c||c|c|}
\hline
\multicolumn{1}{|c|}{$\Enu$} &
\multicolumn{2}{|c||}{$\Fe56$} &
\multicolumn{2}{|c||}{$\Xe132$} &
\multicolumn{2}{|c||}{$\Pb208$}\\
\hline

(MeV)      & $N_n$ &  Sum   & $N_n$ & Sum   & $N_n$   &   Sum   \\
\hline 
0 -- 5    &  0.0   &   0.0   &  0.0   &   0.0   &  0.0    &   0.0   \\
        & (0.0)  &  (0.0)  & (0.0)  &  (0.0)  & (0.0)   &  (0.0)  \\
\hline
 5 -- 10   &  0.0   &   0.0   &  0.01  &   0.01  &  0.0    &   0.0   \\
        & (0.0)  &  (0.0)  & (0.0)  &  (0.0)  & (0.0)   &  (0.0)  \\
\hline
 10 -- 15  &  0.04  &   0.04  &  0.77  &   0.78  &  2.24   &   2.24  \\
        & (0.04) &  (0.04) & (0.84) &  (0.84) & (2.63)  &  (2.63) \\
\hline
 15 -- 20  &  0.23  &   0.27  &  7.38  &   8.16  &  12.44  &  14.68  \\
        & (0.21) &  (0.25) & (6.73) &  (7.57) & (11.91) & (14.54) \\
\hline 
 20 -- 25  &  0.46  &   0.73  &  18.13 &  26.29  &  25.79  &  40.47  \\
        & (0.38) &  (0.63) &(13.80) & (21.37) & (21.06) & (35.60) \\
\hline 
 25 -- 30  &  0.54  &   1.27  &  23.33 &  49.62  &  31.54  &  72.01  \\
        & (0.41) &  (1.04) &(17.62) & (38.99) & (23.59) & (59.19) \\
\hline 
 30 -- 35  &  0.51  &   1.78  &  21.34 &  70.98  &  27.30  &  99.31  \\
        & (0.36) &  (1.40) &(15.36) & (54.35) & (19.23) & (78.42) \\
\hline 
 35 -- 40  &  0.43  &   2.21  &  16.17 &  87.15  &  21.02  & 120.33  \\
        & (0.31) &  (1.71) &(11.37) & (65.72) & (14.34) & (92.76) \\
\hline
 40 -- 45  &  0.35  &   2.56  &  10.87 &  98.02  &  14.74  & 135.07  \\
        & (0.25) &  (1.96) & (7.56) & (73.28) &  (9.98) &(102.74) \\
\hline
 45 -- 50  &  0.27  &   2.83  &   6.74 & 104.76  &   9.38  & 144.45  \\
        & (0.19) &  (2.15) & (4.67) & (77.95) &  (6.15) &(108.89) \\
\hline 
 50 -- 55  &  0.20  &   3.03  &   3.95 & 108.71  &   5.32  & 149.77  \\
        & (0.14) &  (2.29) & (2.73) & (80.68) &  (3.91) &(112.80)  \\
\hline
 55 -- 60  &  0.14  &   3.17  &   2.22 & 110.93  &   3.30  & 153.07  \\
        & (0.10) &  (2.39) & (1.53) & (82.21) &  (2.06) &(114.86) \\
\hline
\end{tabular}
\end{center}
\end{table}

As seen from Figure  
\ref{fig:lum-avEnergy-timeprofile-spectra}, the spectra of the 
neutrinos coming from the SN core have their peaks within 10 MeV and 
then rapidly fall off at higher energies. The neutron numbers have their 
maximum in the interval of 25-30 MeV
and then fall off reaching almost zero by 60 MeV. Though this neutrino
energy dependence cannot be measured experimentally, they indirectly reflect
the energy dependence of the source neutrinos.
 
The total number of emitted neutrons for $\Fe56$, $\Xe132$ and 
$\Pb208$ are 3, 111 and 154, respectively, in the case of NO. Note 
that for $\Pb208$ one neutron comes from neutrino energies 
above 60 MeV and is not seen in Table 
\ref{tab:n-numbers_vs_Enu-bins_nue-CC_NO-IO}. The numbers for the IO 
case for $\Fe56$, $\Xe132$ and $\Pb208$ are 2, 83 and 117, respectively. 
Here again, neutrinos with energies 
above 60 MeV contribute 1 neutron in the case of $\Xe132$ and 2 neutrons 
in the case of $\Pb208$, which are not seen in Table 
\ref{tab:n-numbers_vs_Enu-bins_nue-CC_NO-IO}. The higher 
numbers for NO can be explained by the fact that for NO one has complete 
flavor conversion whereas for IO the conversion is partial. 
\begin{figure}
\begin{center}
\includegraphics[scale=.25, angle=270]{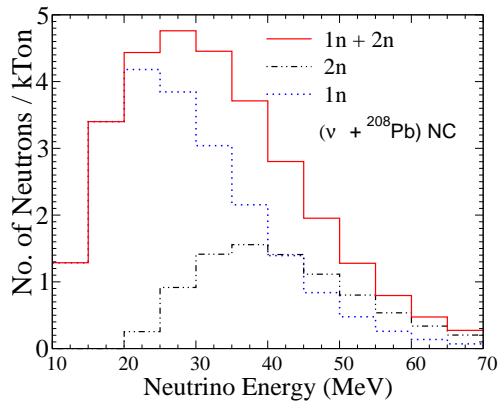}\\
\caption{Number of neutrons emitted as a function of neutrino energy by 
NC excitation of $\Pb208$ for SN neutrino flux given by the
Basel/Darmstadt simulations~\cite{Fischer-10} of a $18\msun$
progenitor supernova at a distance of 10 kpc. 
(From \cite{Bandyopadhyay-17}.)} 
\label{fig:Pb208_histo_n-numbers_vs_Enu_nui-NC}
\end{center}
\end{figure}

The NC excitation of the nuclei is important because here all six neutrino
species contribute. In Figure 
\ref{fig:Pb208_histo_n-numbers_vs_Enu_nui-NC} we present the number of 
neutrons emitted by 1 kton of $\Pb208$ due to NC interaction of all 
neutrino and antineutrino flavors with $\Pb208$. This is given in 
the form of a histogram of 5 MeV energy bins for the neutrino energies. 
The total number of neutrons emitted is 30 with 21 coming from 1 neutron 
emission and 9 from the emission of 2 neutrons. The total number of NC 
neutrons for $\Fe56$ is about 5. Of course, NC gives identical 
numbers for both NO and IO because all neutrino species here contribute 
equally. 

An interesting feature of the above results on the number of neutrons 
produced by different materials is to be noted: As already mentioned in 
Introduction, materials with large neutron excess ($N-Z$) are 
expected to produce 
significantly more number of neutrons through $\nue$ CC interactions 
than through combined NC interactions of all the six neutrino and 
antineutrino species. Thus, for $\Pb208$ with a neutron excess of 44, 
for example, we see that only about 16\% of all neutrons produced come 
from NC interactions and the rest come from CC interaction of the 
$\nue$s. On the other hand, in the case of $\Fe56$ with a 
neutron excess of only 4, more than about 60\% of the produced neutrons 
result from NC interactions of all neutrinos. This offers an 
interesting complementarity between these two detector materials. Thus,  
simultaneous detection of a SN in a lead detector (e.g., the currently 
running HALO detector~\cite{HALO} with about 79 tons of lead) and a 
sufficiently large iron detector (e.g., the proposed ICAL 
detector~\cite{ICAL-INO} with 50 kton of iron, suitably 
instrumented for neutron detection\footnote{The proposed 50 kton 
magnetized Iron Calorimeter (ICAL) detector to be located at the 
proposed India-based Neutrino Observatory (INO) is designed primarily 
for the study of neutrino properties, in particular, the neutrino mass 
hierarchy, using atmospheric neutrinos, and was  
originally not designed to be sensitive to much lower energy SN 
neutrinos. However, it can in principle be modified with suitably 
placed layers of neutron detectors to make the 
detector sensitive to SN neutrino induced neutrons.}) may 
allow one to estimate the fraction of the non-electron flavored 
neutrinos in the SN neutrino flux. This, however, needs a detailed 
analysis that is beyond the scope of this short review.    

\begin{figure}
\begin{center}
\includegraphics[scale=.45]{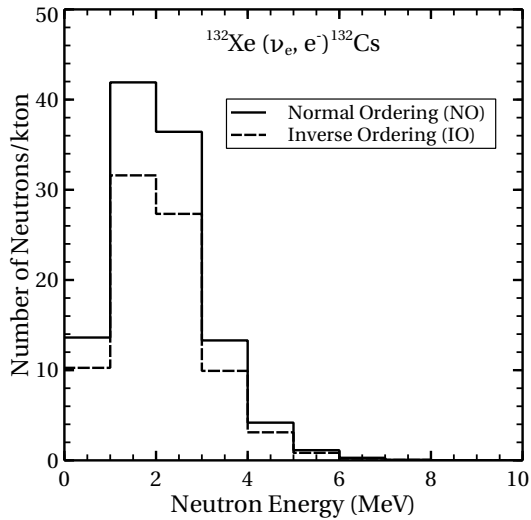}\\
\caption{Spectrum of neutrons emitted by $\Cs132$ excited by SN
neutrinos through the CC process $\Xe132\,(\nue,\eminus)\Cs132$ for both 
NO and IO cases, for SN neutrino flux given by 
the Basel/Darmstadt simulations~\cite{Fischer-10} of a $18\msun$ 
progenitor supernova at a distance of 10 kpc. (From 
\cite{Bandyopadhyay-20}).}
 \label{fig:neutron_spectrum-132xe}
\end{center}
\end{figure}

Finally, it may be possible to experimentally measure the energy 
spectrum of the emitted neutrons. The theoretically calculated spectrum 
turns out to be similar for all the three cases of nuclei with very 
different masses. For illustration we give in Figure 
\ref{fig:neutron_spectrum-132xe} the 
energy spectrum of neutrons emitted by $\Cs132$ due to CC interaction of 
the SN $\nue$s with the nucleus $\Xe132$, for SN neutrino flux given by 
the Basel/Darmstadt simulations~\cite{Fischer-10} of a $18\msun$
progenitor supernova at a distance of 10 kpc. The emitted neutrons are 
seen to reach a peak at around 1.5 MeV and then fall off becoming very 
small beyond 7 MeV or so. 

\section{Summary and outlook}
\label{sec:summary}
In this article we have given a brief review of supernova neutrino 
induced neutron production in different detector materials with primary 
focus on reviewing the nuclear physics aspects 
of neutrino induced nuclear excitation and subsequent de-excitation 
of the nuclei through neutron emission. We have used the 
simulation results of the Basel/Darmstadt group for the explosion of a 
$18\msun$ progenitor star at a distance of 10 kpc for illustrating our 
results for neutron production in different detector materials. 

We have noted that the number of neutrons produced differs for the 
two different neutrino mass orderings, i.e., normal- and inverted 
ordering. While this raises the tantalizing possibility of 
distinguishing between the two mass hierarchies through detection of the 
SN neutrino induced neutrons, this will probably be an extremely 
difficult task given the currently uncertain physics of flavor 
conversion within the supernova and the resulting uncertainties in the 
source spectrum of neutrinos. 

Perhaps more useful from the standpoint of experimental 
exploration is our observation that while the 
dominant fraction of neutrons produced in neutron rich materials such as 
lead comes from charged 
current interaction of the $\nue$s, the opposite happens in a low 
neutron excess material such as iron for which the neutrons produced 
by the combined neutral current interactions of all the six 
neutrino plus antineutrino species dominate. We propose that 
this complementarity between high- and low neutron excess detector 
materials may offer a way of estimating the fraction of mu- and tau 
flavored neutrinos (which interact only through neutral current) in the 
total SN neutrino flux by means of simultaneous detection of a SN 
through the neutron channel in two sufficiently large 
detectors, with one detector made of lead (high neutron excess) and the 
other of iron (low neutron excess), for example. This can give valuable 
information about the production and flavor oscillation processes of 
neutrinos in supernovae. Of course, the amounts of detector material 
that will be needed for each type of detector for drawing statistically 
significant inference on the flavor composition of the SN neutrinos will 
require a detailed analysis that is beyond the scope of the present 
article.              

In the above discussions we have used a specific set of simulation 
results, namely, that  of the Basel/Darmstadt group 
for the explosion of a $18\msun$ progenitor star. It will be 
important to use the results of other simulations that include 
more realistic mean field and modification of the symmetry energy as 
pointed out in recent works, and also consider different 
masses of the progenitor star in order to have a better idea of the 
systematic uncertainties in the prediction of number of neutrons 
produced. Finally, one hopes that a future core 
collapse supernova will be observed at a distance closer than 10 kpc; 
an explosion at a distance of 1 kpc, for example,  
will give 100 times more events than what we have estimated here. 

We thank Abhijit Bandyopadhyay, Sovan Chakraborty and Satyajit Saha for
helpful discussions. One of us (PB) acknowledges support
under a Raja Ramanna Fellowship of the Dept.~of Atomic Energy, Govt.~of 
India. 

\vskip 1.5cm
\noindent{\bf Author contribution statement}\\

\noindent Both the authors have contributed equally to preparation of 
the first draft and subsequent modifications leading to the final 
version of the manuscript.

\end{document}